# Understanding Accessibility Needs of Blind Authors on CMS-Based Websites

Guillermo Vera-Amaro[1,2] [0009-0002-1019-6986] and José Rafael Rojano-Cáceres[1,3] [0000-0002-3878-4571]

[1] Faculty of Statistics and Informatics, University of Veracruz, Xalapa, Veracruz, Mexico
[2] `gvera@uv.mx`
[3] `rrojano@uv.mx`

**Abstract.** This paper addresses the limited attention given to blind users as content creators in Content Management Systems (CMS), a gap that remains underexplored in web accessibility research. For blind authors, effective interaction with CMS platforms requires more than technical compliance; it demands interfaces designed with semantic clarity, predictable navigation, and meaningful feedback for screen reader users. This study investigates the accessibility barriers blind users face when performing key tasks—such as page creation, menu editing, and image publishing—using CMS platforms. A two-fold evaluation was conducted using automated tools and manual usability testing with three blind and one sighted participant, complemented by expert analysis based on the Barrier Walkthrough method. Results showed that block-based interfaces were particularly challenging, often marked as accessible by automated tools but resulting in critical usability issues during manual evaluation. The use of a text-based editor, the integration of AI-generated image descriptions, and training aligned with screen reader workflows, significantly improved usability and autonomy. These findings underscore the limitations of automated assessments and highlight the importance of user-centered design practices. Enhancing CMS accessibility requires consistent navigation structures, reduced reliance on visual interaction patterns, and the integration of AI tools that support blind content authors throughout the content creation process.

**Keywords:** Web Accessibility, Usability Study, User Experience, Artificial Intelligence, Blind Users, Content Management System.

## 1 Introduction

Websites are the primary sources of information and services on the internet [1], and according to the Global Digital Report, around 66% of the world's population has internet access [2], highlighting the need for tools that enable website creation without advanced technical skills [3]. Content management systems (CMS) have become important tools for this purpose, powering over 70% of websites globally [4]. These platforms offer user-friendly interfaces, but many still present significant accessibility barriers for blind users [3], despite efforts to align with WCAG standards [5].

Approximately 16% of the global population—around 1.3 billion people—live with disabilities [6], including visually impaired content creators who face challenges using CMS tools to produce content for both blind and sighted users. In the United States alone, 26% of adults (67 million individuals) have some form of disability [7]. Unfortunately, accessibility is often treated as an afterthought, with some products omitting its implementation altogether [8]. This is partly due to reliance on automated evaluation tools [9], which perform syntactic checks—like verifying alternative text or proper HTML structure—while failing to assess critical semantic aspects such as page hierarchy, element roles, and overall usability. These aspects are essential for effective navigation by screen reader users [8], yet many CMS platforms are not born-accessible [10].

This study evaluates the accessibility of CMS components, focusing on navigation, form interaction, and content management from the perspective of blind authors. Through workshops with blind participants on a dedicated website, we identify usability challenges and assess how effectively a CMS supports accessible content creation. Our main contribution is to document the accessibility needs of blind authors in CMS environments through an empirical approach, emphasizing that key semantic aspects—such as structure, element roles, and functional purposes—are often undetected by automated tools but are critical for a usable and accessible experience.

The remainder of this paper is organized as follows: Section 2 and Section 3 reviews relevant literature on web accessibility and CMS platforms, Section 4 outlines the research methodology, Section 5 presents the key findings, followed by a discussion in Section 6. Limitations of the study are addressed in Section 7, and finally, Section 8 summarizes the conclusions and suggests directions for future research.

## 2    Background

Web accessibility according to the World Wide Web Consortium (W3C) [11], promotes that websites and technologies are usable by people with disabilities. The Web Accessibility Initiative (WAI) [12] plays a key role through standards like the Web Content Accessibility Guidelines (WCAG), while the Authoring Tool Accessibility Guidelines (ATAG) 2.0 [13] promote best practices for making CMS platforms accessible [14].

People with visual impairments require screen readers to access information online. Screen readers agents convert on-screen content into speech [8], relying on semantic HTML and ARIA (Accessible Rich Internet Applications) roles to interpret page structure [15]. Poor semantic markup impairs navigation, creating barriers for visually impaired users. Automated tools can detect basic issues like missing alt text but often fail to evaluate deeper semantic structures, such as proper heading hierarchies [8]. Attempts to improve this include machine learning approaches for identifying ARIA landmarks [16], browser extensions using natural language processing for navigation aids [17], and systems for automatically injecting landmarks [18]. However, consistent with HCI principles on usability and real-world performance [19], true accessibility evaluation requires combining automated assessments with user-based testing [20].

CMS platforms enable content creation without advanced technical skills [21], in that sense WordPress powers 43.6% of all websites globally and  29% of websites do not use a CMS solution [4]. This translates to a content management systems market share of 61%. Despite their flexibility, many CMS platforms present accessibility challenges, such as inconsistent navigation and poor screen reader support [3]. Addressing these issues requires adherence to the WCAG and user-centered design practices [22], including semantic labeling and keyboard navigability [5]. Artificial intelligence tools, like automated image description generators [23], support accessibility by reducing manual effort and improving multimedia engagement for blind users [24].

Accessibility evaluations often combine qualitative (user-centered) and quantitative (automated) approaches [20]. While inspection methods focus on technical compliance, observation methods capture user experiences and real interaction challenges [25]. Prior studies suggest that integrating both approaches provides more comprehensive insights [1]. This research adopts a combined strategy, aligning with HCI principles of usability and user empowerment [19], by integrating automated evaluations with blind user testing to uncover both technical and experiential barriers.

## 3     Related work

We conducted a literature review on accessibility in CMS-based platforms using the search string *("content management systems" OR "wordpress") AND "accessibility"* in the Scopus and Web of Science databases, focusing on barriers encountered by blind users. The search yielded 92 results, among which several relevant studies were identified. Pascual et al. [26], who evaluated Blogger and WordPress, highlighting challenges due to insufficient guidance for accessible design and incomplete adherence to WCAG and ATAG standards, without assessing administrative functions. Similarly, Calvo et al. [27] evaluated Moodle [28], revealing significant barriers for screen reader users in content management workflows.

Regarding content editors, Sanchez-Gordon et al. [22] improved accessibility in TinyMCE and Summernote editors by adding image insertion controls, while López et al. [29] analyzed six CMS platforms, finding that none fully met ATAG guidelines [13]. In a later study, they proposed a methodology to identify and correct accessibility issues [30]. Additionally, Csontos and Heckl et al. [3] explored tools to address issues like missing alt text and poor semantic structure while Acosta et al. [31] defined a method for evaluating the accessibility of online content editors.

CMS-specific solutions include a WordPress plugin by Causarano and Paternò et al. [21], integrated with MAUVE++ for real-time feedback. Zilak et al. [32] also assessed a WordPress prototype using automated evaluations and surveys, though notably without involving screen reader users, limiting real-world insights into usability.

Overall, most studies rely on automated tools for accessibility evaluations, with fewer involving expert reviews and even fewer including blind users. To the best of our knowledge, no previous research has combined automated analysis with usability testing involving blind users as content creators. This study aims to fill that gap, focusing on the barriers they face and providing a foundation for future research.

## 4    Method

This research employs a modified version of the WCAG Evaluation Methodology (WCAG-EM) [33], similar to the approach proposed by Acosta-Vargas et al. [34] with adaptations aligned to our objectives as illustrated in Fig. 1. It integrates quantitative evaluation using automated tools and qualitative analysis through empirical testing with blind users managing CMS content. Manual evaluation also incorporates expert review via the Barrier Walkthrough (BW) method [35].

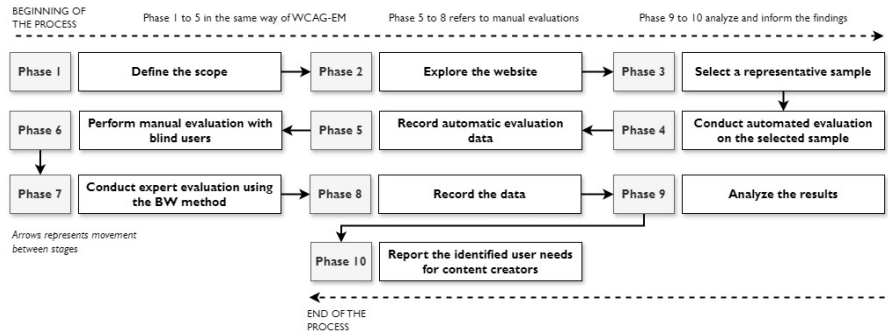

**Fig. 1.** Proposed research method.

### 4.1    Conduction of the study

The method was applied to CMS tasks typically challenging for blind authors, such as those listed in Table 1. Each task was first evaluated using an automated tool, followed by empirical testing across three sessions with blind participants. Barriers identified were rated for severity based on impact on task performance following the BW method.

**Phase 1: Defining the scope.** This study explores the accessibility needs of blind authors, focusing on four research questions:

- **RQ1**. What is the correlation between automated and manual evaluation?
- **RQ2.** How can AI assist in tasks requiring visual input?
- **RQ3.** How does training impact user performance?
- **RQ4.** What challenges do blind authors face in CMS?

WordPress 6.7.1 was deployed on an Ubuntu 24.04.1 virtual machine with MySQL 8.0.41 and Apache 2.4.58. WordPress was selected for its open-source nature, plugin extensibility [36], and alignment with WCAG 2.2 Level AA standards [5], offering features like semantic HTML and keyboard navigation [37, 38]. Training materials specifically designed for screen reader users were developed and published online across three sessions. All sessions were recorded and qualitatively analyzed based on participant feedback.

**Phase 2: Exploring the website.** The CMS interface was analyzed to identify accessibility-relevant functionalities. Built with HTML, CSS, JavaScript, and PHP, it contains dynamic elements that may create barriers for screen readers. Key components include the Toolbar (1) for administrative quick task, the Main Navigation Menu (2) for core feature management, and the Work Area (3) for content creation. Other sections, such as the Page Title (4), Action Notices (5), Filter Area (6), and the Main Content Area (7), are shown in Fig. 2. This exploration highlighted barriers, particularly in complex interaction areas.

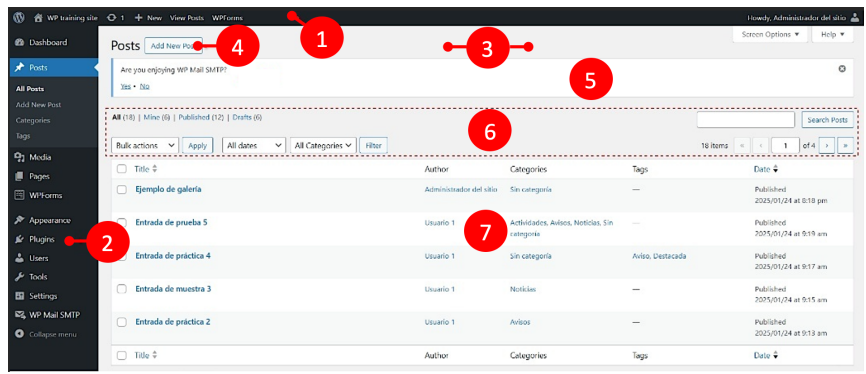

**Fig. 2.** CMS administration interface highlighting key sections.

**Phase 3: Selecting a representative sample.** Since this study focuses on content creation, public-facing website design was not evaluated. A pilot session with a blind user helped define key tasks that blind content authors need to perform in a CMS, following prior research [36], summarized in Table 1.

**Table 1.** Identified tasks that blind content authors need to perform with automated evaluations.

| Id  | Task                              | CMS page         | Errors | Alerts | E+A |
|-----|-----------------------------------|------------------|--------|--------|-----|
| T01 | Login and enter credentials       | wp-login         | 0      | 1      | 1   |
| T02 | Navigate through menus            | wp-admin         | 3      | 17     | 20  |
| T03 | Use block-based editor            | post-new-block   | 2      | 17     | 19  |
| T04 | Use text-based editor             | post-new         | 16     | 23     | 39  |
| T05 | Create and manage categories      | edit-tags        | 14     | 23     | 37  |
| T06 | Manage media files                | upload           | 4      | 18     | 22  |
| T07 | Create image galleries            | post-new-gallery | 19     | 24     | 43  |
| T08 | Configure homepage and blog page  | options-reading  | 2      | 17     | 19  |
| T09 | Adjust general site settings      | options-general  | 2      | 20     | 22  |
| T10 | Edit main menu                    | nav-menus        | 23     | 25     | 48  |
| T11 | Configure social media widget     | widgets          | 2      | 17     | 19  |
| T12 | Customize design                  | customize        | 125    | 118    | 243 |
| T13 | Install and activate complements  | plugins          | 14     | 20     | 34  |
| T14 | Create and manage forms           | form-builder     | 78     | 49     | 127 |

**Phase 4: Automatic evaluation on the selected sample.** WAVE was selected for automated evaluation due to its effectiveness in identifying accessibility issues and WCAG 2.2 support [39]. The WAVE Chrome extension was used for its ease of deployment and visual reporting [14, 40]. Key CMS pages listed in Table 1 were evaluated.

**Phase 5: Record automatic evaluation data.** Accessibility scores from the WAVE evaluation were recorded and organized, and the full dataset[1] is publicly available to ensure transparency and reproducibility.

**Phase 6: Perform the manual evaluation with blind users.** A manual assessment followed, involving five participants: three blind users experienced with screen readers (NVDA and JAWS), a sighted user for visual support, and an instructor with over ten years of CMS experience. Participant demographics are detailed in Table 2.

Table 2. Participants demographic information.

| Id | Gender | Role | Sight Condition | Screen Reader Experience | Screen Reader |
|---|---|---|---|---|---|
| P1 | M | Author | Blind | 18 years | JAWS |
| P2 | M | Author | Partially Blind | 15 years | NVDA |
| P3 | F | Author | Partially Blind | 16 years | NVDA |
| P4 | M | Author | Sighted | - | None |
| P5 | M | Instructor | Sighted | 1 year | NVDA |

A pilot test with one blind participant helped refine the evaluation process, including task duration and accessibility setting and ensuring comfort by allowing participants to use personal keyboards. The manual evaluation included: (1) assessing initial user experience without CMS training, (2) evaluating the same sections as in automated assessment, and (3) conducting three 2-hour training sessions. The first session provided a verbally described overview by the instructor sharing his screen; the second involved the instructor using NVDA, sharing both the screen and audio; and the third focused on complex areas with challenging navigation flows, again using screen reader technology. All sessions were recorded to collect user feedback on their experiences.

**Phase 7: Conduct expert evaluation using the BW method.** Manual evaluation data, gathered from user observation and feedback, were analyzed using an adapted Barrier Walkthrough (BW) method [35]. BW identifies barriers that hinder CMS usability. Observations were converted into quantitative data, following prior studies [1, 32]. Barriers were rated on impact (effect on usability) and persistence (frequency of occurrence). Impact was rated from None (0), Minor (1), Significant (2), to Critical (3), while persistence was based on barrier frequency (See Table 3). The final severity score was determined by combining impact and persistence, following the BW method guidelines. If either impact or persistence was 0, the severity was classified as "None". A "Minor" severity was assigned when impact was 1 and persistence was 1 or 2. A "Significant" severity was given if impact was 2 and persistence was 1 or 2. A "Critical" severity was given if impact was 3 or persistence reached 3 (See Fig. 3). A CMS-specific adaptation of the original barrier list for blind persons was developed, incorporating WCAG accessibility principles [41], which provide the foundation for accessible web content and establish that must be Perceivable, Operable, Understandable, and Robust for effective access and use (See Table 4).

---

[1] All data supporting this study are available at: doi: 10.17632/4f9ptj34nb.3

**Table 3.** Severity rating scale, barrier occurrences and persistence levels.

| Impact | Meaning | Barrier occurrences | Persistence |
|---|---|---|---|
| 0 | None | 0 and 1 | 0 |
| 1 | Minor | 2 and 3 | 1 |
| 2 | Significant | 4 and 5 | 2 |
| 3 | Critical | Greater than 5 | 3 |

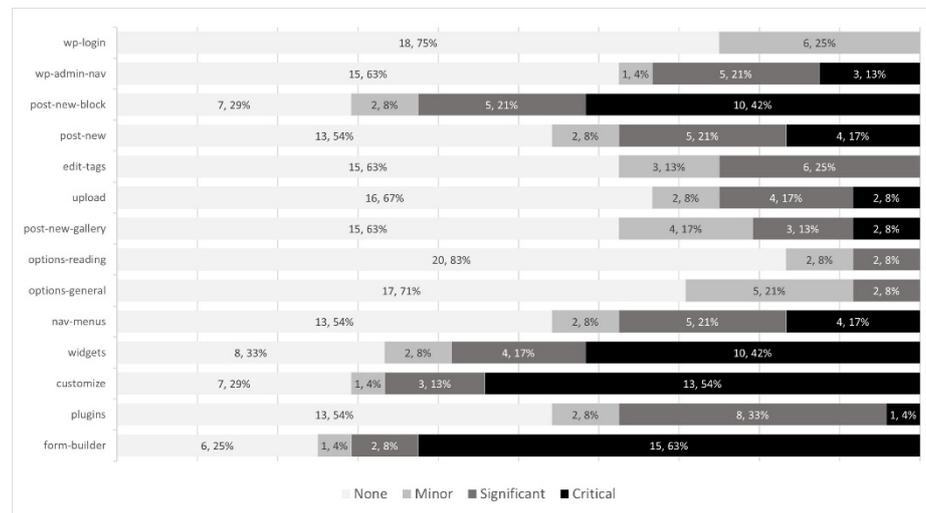

**Fig. 3.** Summary of severity levels for each evaluated CMS task.

**Phase 8: Record the data.** Barrier Walkthrough evaluation data, including impact, persistence, and severity scores, were systematically recorded. Each task evaluation was documented individually and recorded on tables (see Table 4). The complete dataset is available for reference, and summarized results are shown in Table 5.

**Phase 9: Analyze the results.** The lowest-rated area was the block editor, with a Critical severity of 10, followed by Customize Design with a score of 9. The highest-rated task was Login, followed by Set general settings (See Fig. 3). Correlation analysis between automated and manual evaluations was performed (See Fig. 4). These findings, discussed in detail in the Section 5.

Table 4. Sample of evaluation with the BW method for the T03-Use block-based editor.

| Id | Barrier | Principle | Success Criteria | Impact | Persistence | Severity |
|---|---|---|---|---|---|---|
| B01 | Rich images lacking equivalent text | Perceivable | 1.1.1 | 1 | 0 | None |
| B02 | Color is necessary | Perceivable | 1.4.1, 1.4.3 | 1 | 1 | Minor |
| B03 | Opaque objects | Perceivable | 1.1.1 | 1 | 0 | None |
| B04 | Functional images lacking text | Perceivable | 1.1.1, 1.4.5 | 1 | 0 | None |
| B05 | Tables with no structural relationships | Perceivable | 1.3.1 | 0 | 0 | None |
| B06 | Tables with no summary | Perceivable | 1.3.1, 1.3.2 | 0 | 0 | None |
| B07 | Forms that are badly linearized | Perceivable | 1.3.2 | 2 | 0 | None |
| B08 | Generic links | Operable | 2.4.4, 2.4.9 | 1 | 0 | None |
| B09 | Ambiguous links | Operable | 2.4.4, 2.4.9 | 1 | 2 | Minor |
| B10 | Dynamic menus in JavaScript or CSS | Operable | 2.1.1, 2.1.3, 4.1.2 | 0 | 0 | None |
| B11 | Mouse events | Operable | 2.1.1, 2.1.3 | 0 | 0 | None |
| B12 | Keyboard traps | Operable | 2.1.2 | 0 | 0 | None |
| B13 | Too many links | Operable | 2.4.10 | 1 | 1 | Minor |
| B14 | Forms with no LABEL tags | Operable | 3.3.2, 1.3.1 | 2 | 0 | None |
| B15 | Non separated links | Operable | 2.4.4, 2.4.9 | 1 | 0 | None |
| B16 | Skip links not implemented | Operable | 2.4.1 | 1 | 0 | None |
| B17 | No keyboard shortcuts | Operable | 2.1.1, 2.1.4 | 1 | 1 | Minor |
| B18 | No page headings | Operable | 2.4.6, 1.3.1 | 1 | 1 | Minor |
| B19 | Page without titles | Operable | 2.4.2 | 1 | 0 | None |
| B20 | Difficulty locating the save button | Operable | 2.4.3, 1.3.2 | 2 | 0 | None |
| B21 | Form with redirect | Understandable | 3.2,2, 3.2.5 | 1 | 1 | Minor |
| B22 | New windows | Understandable | 3.2.1, 3.2.5 | 2 | 0 | None |
| B23 | Language markup | Understandable | 3.1.1, 3.1.2 | 1 | 0 | None |
| B24 | Dynamic changes | Robust | 4.1.2, 4.1.3 | 2 | 0 | None |

**Phase 10: Report on the content creator's user needs.** A major requirement is consistent and predictable navigation structures, as screen reader users rely on navigation semantics. Visual blocks, dynamic menus, inconsistent heading structures, and ambiguous links created confusion and increased cognitive load.

Table 5. Summary of BW severity levels for all manual evaluated tasks.

| Id | Task | None | Minor | Significant | Critical |
|---|---|---|---|---|---|
| T01 | Login and enter credentials | 18 | 6 | 0 | 0 |
| T02 | Navigate through menus | 15 | 1 | 5 | 3 |
| T03 | Use block-based editor | 7 | 2 | 5 | 10 |
| T04 | Use text-based editor | 13 | 2 | 5 | 4 |
| T05 | Create and manage categories | 15 | 3 | 6 | 0 |
| T06 | Manage media files | 16 | 2 | 4 | 2 |
| T07 | Create image galleries | 15 | 4 | 3 | 2 |
| T08 | Configure homepage and blog page | 20 | 2 | 2 | 0 |
| T09 | Adjust general site settings | 17 | 5 | 2 | 0 |
| T10 | Edit main menu | 13 | 2 | 5 | 4 |
| T11 | Configure social media widget | 8 | 2 | 4 | 10 |
| T12 | Customize design | 7 | 1 | 3 | 13 |
| T13 | Install and activate complements | 13 | 2 | 8 | 1 |
| T14 | Create and manage forms | 6 | 1 | 2 | 15 |

## 5   Results

This section analyzes the findings in response to each of the research questions.

### 5.1  RQ1. What is the correlation between automated and manual evaluation?

To measure the linear relationship between automated evaluation scores and manual user testing severity, the correlation coefficient (1), was applied using the $CORREL()$ función in a spreadsheet software. This coefficient quantifies the strength and direction of the relationship, ranging from $-1$ to $1$. A value of $1$ indicates a perfect positive correlation, greater than $0$ signifies a positive correlation, zero represents no linear relationship, and $-1$ denotes a negative correlation [1].

$$r = \frac{\sum(x-\bar{x})(y-\bar{y})}{\sqrt{\sum(x-\bar{x})^2 \sum(y-\bar{y})^2}} \quad (1)$$

The correlation analysis revealed varying relationships between the automated evaluation scores and manual severity ratings. As shown in Fig. 4, a moderate negative correlation ($r = -0.58$) with "None" severity suggests that as the number of errors and alerts reported by automated tools increases, the likelihood of having no issues decreases. A weak negative correlation ($r = -0.45$) was found with "Minor" severity, while "Significant" severity showed a very weak negative correlation ($r = -0.14$), indicating little to no relationship. Conversely, "Critical" severity showed a moderate positive correlation ($r = 0.69$), suggesting that higher automated error scores align with an increased presence of critical barriers, though some discrepancies remain.

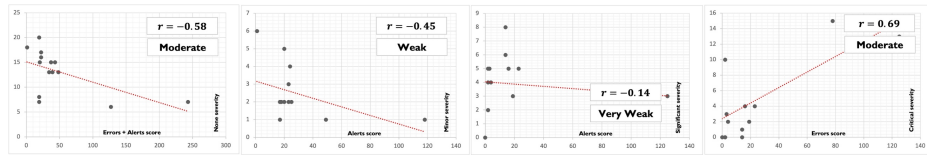

**Fig. 4.** Scatter plots showing the linear correlation between automated and manual evaluations.

### 5.2  RQ2. How can AI assist in tasks requiring visual input?

AI-based tools proved valuable in addressing specific accessibility barriers, particularly those related to visual content lacking alternative text [42]. This directly aligns with WCAG 2.2 criterion 1.1.1 (Non-text Content), which although not one of the most frequently violated overall, becomes critically impactful when associated with functional images (Barrier B04), as confirmed by multiple severe cases in our evaluation. During publishing tasks, blind authors often uploaded images without alt text, making them invisible to screen reader users. The integration of AI-powered plugins into WordPress enabled the automatic generation of descriptive alt text, reducing the need for sighted assistance and supporting more independent content creation workflows.

### 5.3     RQ3. How does training impact user performance?

Before training, blind participants struggled with interface navigation, especially in block-based layouts, often encountering keyboard traps and relying on inefficient strategies such as excessive tabbing, as reported in [18]. These issues relate to WCAG criteria 2.1.2 (No Keyboard Trap) and 2.4.3 (Focus Order), as well as Nielsen's heuristics on user control and visibility of system status [19]. After targeted instruction, user performance improved significantly. Participants navigated more efficiently, used headings effectively, and completed tasks with greater success. The second session, where the instructor used NVDA alongside participants, was particularly impactful. Hearing synchronized feedback from both the instructor's and their own screen reader helped participants better understand navigation sequences, reducing confusion and enhancing learnability and efficiency.

Training emphasized screen reader commands and semantic navigation strategies, enabling users to better identify form fields (WCAG 1.3.1), operate menus, and publish content while reducing errors—supporting both error prevention and help and documentation heuristics. The use of assistive technology in training created a shared learning environment that eased cognitive load and allowed users to follow instructions more intuitively, confirming that training methods aligned with assistive workflows can significantly improve usability for blind content creators.

### 5.4     RQ4. What challenges do blind authors face in CMS?

Blind content authors encounter substantial usability challenges in CMS environments due to poor semantic structure, lack of consistent feedback, and visual interaction patterns. Automated evaluation revealed a median of 14 errors and 22 alerts per page, while manual testing identified 4 critical, 5 significant, and 2 minor severities. The most critical barriers included too many links (B13) and difficulty locating the save button (B20), reflecting violations of principles related to efficiency and error prevention. These issues disrupt the user's mental model and increase the time and effort required to complete tasks.

From a Human-Computer Interaction (HCI) perspective, poor affordances in block-based editors—relying on drag-and-drop interactions [43]—hinder learnability and lead to "dead ends" where navigation fails. Barriers such as ambiguous links (B09), dynamic menus without ARIA roles (B10), and keyboard traps (B12) correspond to frequent violations of WCAG criteria 2.4.4, 2.1.1, and 4.1.2. These directly impact operability and reduce user control, requiring blind users to rely on inefficient strategies like excessive tabbing. In contrast, criteria such as 1.1.1 (alt text) and 3.1.1/3.1.2 (language identification) showed lower severity, indicating better baseline compliance.

A notable cognitive burden stems from insufficient or inaccessible status messages, tied to criterion 4.1.3 (B24), which limits feedback during content creation. This lack of immediate system response violates HCI heuristics related to visibility of system status. Similarly, excessive or irrelevant link groups (B13) reduce navigational clarity, increasing error likelihood. Although tools powered by AI help generate alt text [44], these must be fully integrated into the CMS to support blind authors effectively and reduce visual dependency. Classic text-based editors like TinyMCE yielded better performance due to their predictable layout and keyboard compatibility, enhancing efficiency and user satisfaction. These findings emphasize the need for CMS platforms to move beyond visual paradigms and adopt inclusive design practices that align with both WCAG compliance and core HCI principles.

## 6    Discussion

Automated tools like WAVE can identify basic accessibility issues but fall short in assessing real usability and user experience [14]. This was evident in the evaluation of the block-based editor, which was marked as accessible yet proved highly problematic during manual testing. Critical barriers were significantly more frequent than minor or significant ones, especially under the Robust principle—often due to inaccessible dynamic changes and lack of appropriate feedback (WCAG 4.1.3). The Operable principle also showed notable violations related to navigation and keyboard interactions (WCAG 2.1.1, 2.4.4). In contrast, Understandable criteria—such as labels and language identification (WCAG 3.1.1, 3.1.2)—were less frequently violated. As a result, some CMS pages offer moderate accessibility, whereas others present significant challenges (See Fig. 5). Pages with simpler, sequential structures (e.g., login or general settings) aligned better across automated and manual evaluations. However, visual interfaces such as the Customizer and Form Builder consistently scored poorly, indicating that layout complexity and visual-centric design increase cognitive load and hinder usability. These findings highlight the need for learnable interfaces that provide consistent navigation models and reduce ambiguity in user flows—especially for blind users relying on screen readers.

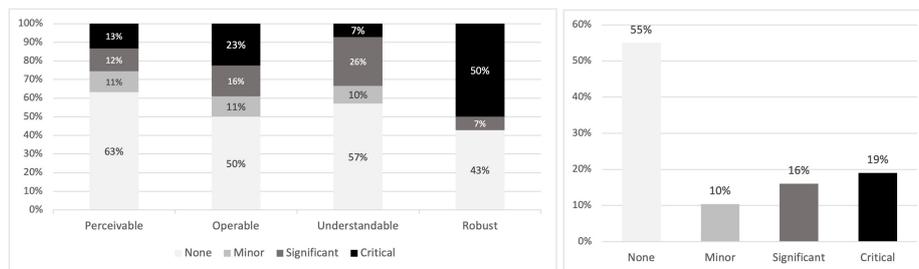

**Fig. 5.** Distribution of severity levels of identified barriers across accessibility principles.

From a HCI lens, the training modality had a strong impact on user performance. Sessions where the instructor used NVDA improved learnability and task efficiency, reinforcing the importance of synchronized feedback and multimodal instruction. Furthermore, the use of AI-powered tools to generate alt text supported error prevention and user autonomy, enabling blind authors to produce accessible content with less reliance on sighted support. However, these improvements reveal a deeper issue: usability barriers often stem not from user limitations, but from systems that overlook inclusive design principles [45].

CMS development must evolve beyond checklist compliance and embed user-centered design into its development lifecycle. This involves involving disabled users in evaluations, integrating manual testing, and ensuring that design decisions address real-world challenges faced by blind content authors—not only to meet accessibility standards, but to uphold digital inclusion as a fundamental right.

### 6.1     HCI design recommendations for accessible CMS

From a user-centered design perspective, the findings suggest concrete improvements for CMS developers and theme/plugin designers. First, CMS interfaces should avoid drag-and-drop mechanics that exclude screen reader users and instead prioritize keyboard operability and semantic clarity (WCAG 2.1.1, 1.3.1). Dynamic components, such as expandable menus or modals, must be properly labeled with ARIA roles to ensure clarity and feedback (WCAG 4.1.2, 4.1.3). Navigation structures should be predictable and support heading-based movement (WCAG 2.4.1, 2.4.6), minimizing cognitive load. Additionally, form elements must include visible labels and provide status confirmation messages (WCAG 3.3.2), reducing errors and supporting learnability. Finally, integrating AI-driven tools to assist with tasks such as image description aligns with the autonomy needs of blind users and should be offered as a native option, not just as third-party extensions.

## 7     Limitations

This study has some limitations that may affect result generalizability. The small sample size of four participants could be considered a constraint; however, it adheres to established protocols recommending a maximum of five visually impaired users to ensure consistent feedback [25, 46]. A pilot test was also conducted with one blind user to refine the procedure. Another limitation involves the BW method, which, while providing valuable insights, depends on expert judgment and can introduce bias. To address this, participant feedback was used to validate findings. Additionally, only one automated tool was used—intentionally—to maintain consistency and evaluate its correlation with manual testing rather than tool comparison. Finally, since the study was conducted remotely, direct observation was limited. This was mitigated by encouraging participants to share their screens during the pilot and describe their navigation processes, helping identify consistent patterns and enhancing result reliability.

## 8     Conclusions and future work

This study examined the accessibility barriers faced by blind content creators when using CMS platforms. Through the combined application of the BW method [35] and the WCAG-EM [33], we identified critical usability issues related to navigation consistency, semantic clarity, and content management workflows. The findings reinforce that blind authors often struggle not because of personal limitations, but due to systemic design choices that prioritize visual interaction. Tailored training, particularly aligned with screen reader workflows, proved transformative, significantly improving users' efficiency and autonomy.

Future work could extend this research by integrating empirical approaches like BW with multiple automated evaluation tools, allowing for a more comprehensive analysis across various user profiles and disabilities. Comparative studies involving other CMS platforms could uncover more accessible solutions or inspire improvements to WordPress, especially regarding block-based interfaces that currently present severe challenges. Expanding user testing to include low vision, motor-impaired, and deaf users would also offer a broader perspective on inclusive CMS design.

Ultimately, addressing the accessibility needs of blind content creators transcends technical enhancements—it aligns with the broader imperative of recognizing accessibility as a fundamental digital right. Ensuring that CMS platforms are operable, perceivable, understandable, and robust is not optional, but essential to fostering equitable participation in digital spaces. Accessible content creation environments empower blind authors to contribute fully and independently to the web, supporting the core principles of universal design and human-centered computing.

**Acknowledgments.** This work is supported by SECIHTI under scholarship CVU: 1351657.

**Disclosure of Interests.** The authors have no competing interests to declare that are relevant to the content of this article.

# References

1. Acosta-Vargas, P., Antonio Salvador-Ullauri, L., Lujan-Mora, S.: A Heuristic Method to Evaluate Web Accessibility for Users with Low Vision. IEEE Access. 7, (2019). https://doi.org/10.1109/ACCESS.2019.2939068.
2. Kemp, S.: Digital 2024: Global Overview Report, https://datareportal.com/reports/digital-2024-global-overview-report, (2024), last accessed 2025/01/29.
3. Csontos, B., Heckl, I.: Improving accessibility of CMS-based websites using automated methods. Univers Access Inf Soc. 21, 491–505 (2022). https://doi.org/10.1007/s10209-020-00784-x.
4. W3Techs: Usage Statistics and Market Share of Content Management Systems, https://w3techs.com/technologies/overview/content_management, last accessed 2025/01/26.
5. WordPress Foundation: Accessibility Coding Standards, https://developer.wordpress.org/coding-standards/wordpress-coding-standards/accessibility/, last accessed 2025/01/26.
6. WHO: Global report on health equity for persons with disabilities. World Health Organization, https://www.who.int/publications/i/item/9789240063600, last accessed 2024/06/10.
7. CDC: Disability and Health Overview, https://www.cdc.gov/ncbddd/disabilityandhealth/disability.html, (2024).
8. Bajammal, M., Mesbah, A.: Semantic web accessibility testing via hierarchical visual analysis. Proceedings - International Conference on Software Engineering. 1610–1621 (2021). https://doi.org/10.1109/ICSE43902.2021.00143.
9. Ismailova, R., Inal, Y.: Comparison of Online Accessibility Evaluation Tools: An Analysis of Tool Effectiveness. IEEE Access. 10, (2022). https://doi.org/10.1109/ACCESS.2022.3179375.
10. Lazar, J.: A Framework for Born-Accessible Development of Software and Digital Content. In: Lecture Notes in Computer Science (including subseries Lecture Notes in Artificial Intelligence and Lecture Notes in Bioinformatics). pp. 333–338. Springer, Cham (2023). https://doi.org/10.1007/978-3-031-42293-5_32.

11. Henry, S.L.: WCAG 2 Overview, https://www.w3.org/WAI/standards-guidelines/wcag/, (2024).
12. W3C WAI: About W3C WAI, https://www.w3.org/WAI/about/, last accessed 2024/09/21.
13. Henry, S.L.: Authoring Tool Accessibility Guidelines (ATAG) Overview, https://www.w3.org/WAI/standards-guidelines/atag/, (2023).
14. Acosta-Vargas, P., Acosta, T., Lujan-Mora, S.: Challenges to assess accessibility in higher education websites: A comparative study of Latin america universities. IEEE Access. 6, (2018). https://doi.org/10.1109/ACCESS.2018.2848978.
15. Nurthen, J., Cooper, M., Henry, S.L.: WAI-ARIA Overview, https://www.w3.org/WAI/standards-guidelines/aria/, (2024).
16. Watanabe, W.M., de Lemos, G., Nascimento, R.W.: Accessibility landmarks identification in web applications based on DOM elements classification. Univers Access Inf Soc. 23, 765–777 (2024). https://doi.org/10.1007/s10209-022-00959-8.
17. Silva, J.S.R., Cardoso, P.C.F., De Bettio, R.W., Tavares, D.C., Silva, C.A., Watanabe, W.M., Freire, A.P.: In-Page Navigation Aids for Screen-Reader Users with Automatic Topicalisation and Labelling. ACM Trans Access Comput. 17, 1–45 (2024). https://doi.org/10.1145/3649223.
18. Aydin, A.S., Feiz, S., Ashok, V., Ramakrishnan, I.: SaIL: Saliency-Driven Injection of ARIA Landmarks. In: Proceedings of the 25th International Conference on Intelligent User Interfaces. pp. 111–115. ACM, New York, NY, USA (2020). https://doi.org/10.1145/3377325.3377540.
19. Nielsen, J.: 10 Usability Heuristics for User Interface Design, https://www.nngroup.com/articles/ten-usability-heuristics/, last accessed 2024/11/02.
20. Luján-Mora, S., Masri, F.: Evaluation of Web Accessibility: A Combined Method. In: Information Systems Research and Exploring Social Artifacts. pp. 314–331. IGI Global (2013). https://doi.org/10.4018/978-1-4666-2491-7.ch016.
21. Causarano, G., Paternò, F.: Facilitating Development of Accessible Web Applications. In: Human-Centered Software Engineering. pp. 47–62. Springer, Cham (2024). https://doi.org/10.1007/978-3-031-64576-1_3.
22. Sanchez-Gordon, S., Calle-Jimenez, T., Villarroel-Ramos, J., Jadán-Guerrero, J., Guevara, C., Lara-Alvarez, P., Acosta-Vargas, P., Salvador-Ullauri, L.: Implementation of Controls for Insertion of Accessible Images in Open Online Editors Based on WCAG Guidelines. Case Studies: TinyMCE and Summernote. In: Advances in Intelligent Systems and Computing. pp. 315–326. Springer, Cham (2019). https://doi.org/10.1007/978-3-030-20040-4_29.
23. Tiwary, T., Mahapatra, R.P.: Web Accessibility Challenges for Disabled and Generation of Alt Text for Images in Websites using Artificial Intelligence. 2022 3rd International Conference on Issues and Challenges in Intelligent Computing Techniques, ICICT 2022. (2022). https://doi.org/10.1109/ICICT55121.2022.10064545.
24. Yadav, A.R., Yadav, S.S., Kamoji, S.: Artificial Intelligence Enhanced Content Management Systems: Integration, Considerations, and Useful Examples. Proceedings - 2023 3rd International Conference on Pervasive Computing and Social Networking, ICPCSN 2023. 283–288 (2023). https://doi.org/10.1109/ICPCSN58827.2023.00053.
25. Ferreira, S.B.L., Da Silveira, D.S., Capra, E.P., Ferreira, A.O.: Protocols for Evaluation of Site Accessibility with the Participation of Blind Users. Procedia Comput Sci. 14, 47–55 (2012). https://doi.org/10.1016/J.PROCS.2012.10.006.
26. Pascual, A., Ribera, M., Granollers, T.: Perception of accessibility errors to raise awareness among Web 2.0 users. ACM International Conference Proceeding Series. (2012). https://doi.org/10.1145/2379636.2379652.
27. Calvo, R., Iglesias, A., Moreno, L.: Accessibility barriers for users of screen readers in the Moodle learning content management system. Univers Access Inf Soc. 13, 315–327 (2014). https://doi.org/10.1007/S10209-013-0314-3.
28. Dougiamas, M.: The Moodle Story, https://moodle.com/about/the-moodle-story/, last accessed 2025/02/02.
29. López, J.M., Pascual, A., Masip, L., Granollers, T., Cardet, X.: Influence of Web Content Management Systems in Web Content Accessibility. In: Human-Computer Interaction. pp. 548–551. Springer, Berlin, Heidelberg (2011). https://doi.org/10.1007/978-3-642-23768-3_79.

30. López, J.M., Pascual, A., Menduĩa, C., Granollers, T.: Methodology for identifying and solving accessibility related issues in web content management system environments. W4A 2012 - International Cross-Disciplinary Conference on Web Accessibility. (2012). https://doi.org/10.1145/2207016.2207043.
31. Acosta, T., Acosta-Vargas, P., Salvador-Ullauri, L., Luján-Mora, S.: Method for Accessibility Assessment of Online Content Editors. In: Advances in Intelligent Systems and Computing. pp. 538–551. Springer, Cham (2018). https://doi.org/10.1007/978-3-319-73450-7_51.
32. Zilak, M., Keselj, A., Besjedica, T.: Accessible Web Prototype Features from Technological Point of View. In: 2019 42nd International Convention on Information and Communication Technology, Electronics and Microelectronics (MIPRO). pp. 457–462. IEEE (2019). https://doi.org/10.23919/MIPRO.2019.8757115.
33. Henry, S.L., Abou-Zahra, S.: WCAG-EM Overview: Website Accessibility Conformance Evaluation Methodology, https://www.w3.org/WAI/test-evaluate/conformance/wcag-em/, (2020).
34. Acosta-Vargas, P., Salvador-Ullauri, L., Pérez-Medina, J.L., Gonzalez, M., Jimenes, K., Rybarczyk, Y.: Improving Web Accessibility: Evaluation and Analysis of a Telerehabilitation Platform for Hip Arthroplasty Patients. Advances in Intelligent Systems and Computing. 959, 508–519 (2020). https://doi.org/10.1007/978-3-030-20040-4_46.
35. Brajnik, G.: The barrier walkthrough method. (2011).
36. Avila, J., Sostmann, K., Breckwoldt, J., Peters, H.: Evaluation of the free, open source software WordPress as electronic portfolio system in undergraduate medical education. BMC Med Educ. 16, 1–10 (2016). https://doi.org/10.1186/S12909-016-0678-1/FIGURES/3.
37. WordPress Foundation: WordPress Accessibility Handbook, https://make.wordpress.org/accessibility/handbook/, last accessed 2025/01/26.
38. Serra, G., Fara, P., Casini, D.: Enhancing the Availability of Web Services in the IoT-to-Edge-to-Cloud Compute Continuum: A WordPress Case Study. Proceedings - 2023 26th Euromicro Conference on Digital System Design, DSD 2023. 602–609 (2023). https://doi.org/10.1109/DSD60849.2023.00088.
39. WebAIM: About WAVE, https://wave.webaim.org/about, last accessed 2025/01/30.
40. Manca, M., Palumbo, V., Paternò, F., Santoro, C.: The Transparency of Automatic Web Accessibility Evaluation Tools: Design Criteria, State of the Art, and User Perception. ACM Trans Access Comput. 16, (2023). https://doi.org/10.1145/3556979.
41. Campbell, A., Adams, C., Bradley, R.: Introduction to Understanding WCAG 2.2, https://www.w3.org/WAI/WCAG22/Understanding/intro#understanding-the-four-principles-of-accessibility, last accessed 2025/02/21.
42. Vera-Amaro, G., Rojano-Cáceres, J.R.: Towards accessible website design through artificial intelligence: A systematic literature review. Inf Softw Technol. 186, 107821 (2025). https://doi.org/10.1016/j.infsof.2025.107821.
43. WP Accessibility Team: Accessibility tickets, tasks, and reports, https://make.wordpress.org/accessibility/handbook/get-involved/tickets-tasks-reports/, last accessed 2025/01/26.
44. Leotta, M., Mori, F., Ribaudo, M.: Evaluating the effectiveness of automatic image captioning for web accessibility. Univers Access Inf Soc. 22, 1293–1313 (2023). https://doi.org/10.1007/s10209-022-00906-7.
45. Chundury, P., Thakkar, U., Reyazuddin, Y., Jordan, J.B., Elmqvist, N., Lazar, J.: Understanding the Visualization and Analytics Needs of Blind and Low-Vision Professionals. In: The 26th International ACM SIGACCESS Conference on Computers and Accessibility. pp. 1–5. ACM, New York, NY, USA (2024). https://doi.org/10.1145/3663548.3688496.
46. Nielsen, Jakob.: Why You Only Need to Test with 5 Users, https://www.nngroup.com/articles/why-you-only-need-to-test-with-5-users/, last accessed 2024/11/02.